\documentclass[twocolumn,prb,showpacs]{revtex4}

\usepackage{graphicx}
\usepackage{rotating}
\usepackage{amsmath}
\usepackage{amsfonts}
\usepackage{amssymb}
\usepackage{enumerate}
\usepackage{longtable}
\usepackage{dcolumn}
\usepackage{bm}

\begin{document}

\newcommand{\be}{\begin{equation}}
\newcommand{\ee}{\end{equation}}
\newcommand{\bn}{\begin{eqnarray}}
\newcommand{\en}{\end{eqnarray}}
\newcommand{\bq}{\begin{eqnarray}}
\newcommand{\eq}{\end{eqnarray}}

\title{Effect of Short-Range Fluctuations on Thermodynamic and 
Resistive Properties: The case of Ising Order}

\author{Mukul S. Laad}
\email[Corresponding author; mukul@physik.rwth-aachen.de]{}
\affiliation{Institut 
f\"ur Theoretische Physik, RWTH Aachen University, 52056 Aachen, Germany}

\author{Luis Craco}
\affiliation{Technische Universit\"at Dresden, 
Institut f\"ur Physikalische Chemie und Elektrochemie, 
01062 Dresden, Germany}

\date{\today}

\begin{abstract}
We consider the effects of the non-local Ising-like ``core spin'' 
correlations on the order-parameter fluctuation contribution to the 
resistivity and thermodynamics of metals showing Ising-like order at finite 
temperature.  We employ the well-known cluster-variation method, and present
explicit results in the pair approximation for short-range order. Our 
calculation generalizes earlier works, where such effects were considered 
in the mean-field (Ornstein-Zernicke) approximation. The mean-field (MF) 
transition temperature $T_{c}^{MF}$, is corrected to $O(1/d)$, and the 
effect of the Ising spin fluctuations on the $dc$ resistivity and 
magnetothermal responses is analyzed in detail. The method can be 
applied straightforwardly to lattices in arbitrary $d$, and, as an 
appealing feature, it reproduces the exact correlation length and 
$T_{c}^{1d}=0$ for the $1d$ Ising model. We apply our results for 
two interesting physical cases: (i) the double-exchange model with 
$J_{H}>>t$, where the core-spins can be approximated quite well by 
Ising spins, and (ii) a model of band electrons coupled to a 
{\it localized} subsystem which undergoes a nematic ordering 
transition coupled to an appropriate structural transition.               
\end{abstract}


\maketitle

\section{INTRODUCTION}

Finite temperature ($T$) order in solid state systems leave their imprint 
on thermodynamic and transport responses across the order-disorder transition.
The steady enhancement of the correlation length of the system as $T$ is 
lowered toward $T_{c}$ from above directly impinges, in particular, on the 
$T$-dependent carrier scattering rate, and manifests itself as changes in
thermal and transport behavior as the phase transition is approached.  For a 
classical transition from a ferromagnetic metal to a paramagnet, such effects
on the $dc$ resistivity for the Heisenberg universality class were considered 
more than forty years ago~\cite{langer}.  This estimation has 
been invaluable in interpreting a large body of data on band ferromagnets. 

Many real compounds, however, show pronounced effects of {\it short-range 
order} (SRO) on carrier properties.  Devising computational schemes capable 
of incorporating such SRO effects into realistic microscopic calculations 
has long been the goal in the field of electronic structure. However, to 
date, it has proved to be extremely demanding to go beyond the single-site 
approximation: in disordered transition-metal alloys, the coherent potential 
approximation (CPA) has long been used with great success~\cite{gyorffy}, 
while for strongly correlated electronic systems, dynamical-mean-field-theory 
(DMFT) in conjunction with {\it ab initio} (LDA) band structure also holds 
a lot of promise~\cite{kotliar}. However, by construction, such a single-site 
approach {\it cannot} capture the full effects of SRO on carrier scattering 
rates.  This requires extending DMFT(CPA) to capture SRO effects via proper 
cluster extensions, a formidable task realizable currently only for 
one-orbital models~\cite{kotliar}.  Moreover, details of such SRO induced 
effects on carrier scattering rates and transport properties remain 
unexplored avenues for study, to our best knowledge.

Motivated by the above, we focus on a much ``simpler'' version of the above 
problem that nevertheless bears on a number of interesting issues in modern
condensed matter physics.  Specifically, we consider physical systems with
an Ising like phase transition to an ordered phase (either rigorously or to 
a very good approximation).  It turns out that, for this particular case, 
a controlled estimation of corresponding order parameter fluctuations 
to $O(1/d)$ (with $d$ the system dimensionality) and their feedback on 
band-like carriers can indeed be carried out under restricted conditions 
(see below).  

To emphasize the relevance of our ``simple'' theory for physical problems
of great current interest, we consider two explicit cases: (i) the double 
exchange (DE) model with $J_{H}>>t$, in which case, the spin correlations
can be well approximated by an {\it effective} Ising model.  This is
believed to be an appropriate low-energy model for certain colossal
magnetoresistance manganites in a limited doping region corresponding to a
well-ordered low $T$ ferromagnet~\cite{furukawa}. (ii) the problem of an 
electronic system undergoing an electronic nematic transition~\cite{carlson}, 
coupled, by symmetry, to an appropriate lattice distortion.  This ordering 
{\it rigorously} belongs to the universality class of an Ising model in an 
external zeeman field.  Evidence for electronic nematicity appears in
various cases of great current interest, like bilayer $Sr_{3}Ru_{2}O_{7}$, 
underdoped Fe arsenides and underdoped high-$T_{c}$ cuprates.
    
\section{FORMALISM}

\subsection{Boundaries of Validity}

In this section, we describe the well-known cluster variation method (CVM)
for the simple DE model ((i) above).  But obviously, the formalism applies 
equally well for case (ii) with replacement of the real Ising spin by a 
$S=1/2$ pseudospin corresponding to Ising nematic order.
 
We start with the double exchange (DE) model in the limit $J_{H}>>t$
with a nearest neighbor ferromagnetic (FM) coupling between core 
spins~\cite{furukawa}. This model is believed to be applicable to 
doped Europium oxide (EuO) and is an effective low-energy model for 
well-doped manganites.  It reads
\be
H=\sum_{<ij>\sigma}t_{ij}({\bf S})(c_{i\sigma}^{\dag} c_{j\sigma}+h.c) -
J'\sum_{<ij>}S_{i}^{z}S_{j}^{z} - h_{ext}\sum_{i}S_{i}^{z} \;.
\ee

In the above, we have included only the Ising part of the intersite FM 
coupling, because the transverse spin fluctuations make a negligible 
contribution when $J_{H}>>t$.  In this limit, the spin correlations are well
approximated by that of an effective Ising model, with an effective intersite 
coupling $J'=(t^{2}/J_{H})x(1-x)$ for the well-doped FM metal.  Also, 
$t_{ij}({\bf S})=t[1+<S_{i}^{z}S_{j}^{z}>/2S^{2}]^{1/2}$ and $h_{ext}$ is the 
external magnetic field.  We emphasize that $t_{ij}$ is to be understood as a 
renormalized carrier hopping integral, reduced from its bare band structure
value by strong local Hubbard correlations in the real system.

In the case of the nematic order, the ``core spin'' above is replaced by 
a nematic pseudospin, $N^{z}$, which quantifies the degree of {\it electronic}
anisotropy, reflecting spontaneous breaking of (discrete) lattice rotational
symmetry~\cite{carlson}. In an Ising nematic, this is defined simply as
$N_{i}^{z}=\frac{n_{i,x}-n_{i,y}}{n_{i,x}+n_{i,y}}$, where $n_{a}$ is the average
fermion occupation number along $a(=x,y)$.  Physically, this can arise from 
multi-orbital effects (as probably the case in $Sr_{3}Ru_{2}O_{7}$~\cite{arun} 
and underdoped Fe arsenides~\cite{phillips}) or from anisotropic (effective) 
two-body interactions~\cite{hae} (as possibly in underdoped cuprates).  
We will {\it assume} that electronic nematic (e-nematic) order has occured, 
without specifying its microscopic origin in this work. By {\it symmetry}, 
$N^{z}$ directly couples to the lattice {\it strain}, which acts as a field 
term conjugate to the order parameter.  The Hamiltonian is
\be
H=\sum_{<ij>\sigma}t_{ij}(c_{i\sigma}^{\dag} c_{j\sigma}+h.c) -
J'\sum_{<ij>}N_{i}^{z}N_{j}^{z} - \delta(T)\sum_{i}N_{i}^{z} \;,
\ee
where $\delta(T)$ is the lattice strain, simplified here to a single 
$T$-dependent function to illustrate its dominant role.  The intersite 
exchange is ``ferro'', since the e-nematic state involves a ${\bf q}=0$ 
instability. Again, $t_{ij}$ is to be understood as an {\it effective} 
hopping of (substantially) renormalised band carriers in reality.
   
Effects of arbitrary SRO on carrier dynamics in real materials has hitherto 
relied on heavy-duty numerical approaches~\cite{elbio,pinaki}. Ideally, one 
would like to extend the local approximation in a way that is amenable to 
treatment of the problem in any dimension, and is computationally tractable. 
In what follows, we present an analytical version of the cluster-variation 
method (CVM), which has been very widely used in studies of alloy formation 
and stability. This scheme satisfies the above requirements,is physically 
transparent and amenable to easy numerical implementation.  We emphasize 
that, as a first step, we look at these effects within effective models, 
and our approach is only valid in the cases where SRO arises from 
(electronic) degrees of freedom that are {\it effectively} decoupled from the 
band-like carriers at low energy.  This is also the reason why we cannot 
describe the microscopics of the ordering mechanisms: we take this as given.

\subsection{Cluster Variation Method}

As a first step, we realize that the core (Ising) spins can be thought of as
an alloy system of $S^{z}=-1/2,+1/2$.  A perfect FM order (low $T$) corresponds
to a pure system of either $S^{z}=-1/2$ or $+1/2$.  At sufficiently high $T$,
the system is non-magnetic and corresponds to a completely random alloy of
$S^{z}=-1/2,+1/2$, where the CPA should be a very good approximation.  At
intermediate $T$, however, SRO effects are dominant.  In this situation, we 
map our problem to that of a short-range ordered binary alloy, which has
been extensively studied in the field of magnetism of disordered transition 
metals.  In particular, we employ the CVM~\cite{fontaine}, which has been 
very successfully used in this context.  The CVM gives analytical expressions 
for the correlation functions of the system within the 
so-called pair and square approximations~\cite{sanchez}.  In addition, it 
satisfies the "diffuse intensity sum rule" exactly in $d=1$, a feature that
 is {\it not} shared by other, less sophisticated approximations.  We make the
identification $S_{i}^{z}=-1/2 \rightarrow n_{i}=0$ and $S_{i}^{z}=1/2 
\rightarrow n_{i}=1$. We start by writing down the functional:
\be
F[\sigma]=\sum_{\sigma}E(\sigma)X(\sigma)
+k_{B}T\sum_{\sigma}X(\sigma)lnX(\sigma) \;,
\ee
where (for a binary system such as the one we consider) for $N$ lattice sites,
the sum runs over $2^{N}$ configurations.  Here, $E(\sigma)$ and $X(\sigma)$
denote the energy and probability of a configuration ${\sigma}$.  The system
free energy is just the minimum of $F[\sigma]$, where the minimization is 
carried out over all density matrices $X(\sigma)$ subject to normalization:
$\sum_{\sigma}X(\sigma)=1$.

In general, the average configurational energy of the spin subsystem can be
written down as a linear combination of many-body correlations.  For a binary
system,
\be
E(\sigma)=1/2\sum_{ij}J_{j}'S_{i}^{z}S_{j}^{z} + \sum_{i}h_{i}S_{i}^{z} \;,
\ee
and
\bn
\nonumber
F[\sigma] &=& E(\sigma)+\frac{k_{B}T}{2}\sum_{ij\sigma\sigma'}
y_{\sigma\sigma'}(i,j) ln y_{\sigma\sigma'}(i,j) \\
&+& k_{B}T(1-z)\sum_{i,\sigma}x_{\sigma}(i) ln x_{\sigma}(i)\;,
\en
where $x_{\sigma}(i)=[1+\sigma\xi_{1}(i)]$ is the single-site occupation
probability and $y_{\sigma\sigma'}(i,j)=(1/4)[1+\sigma\xi_{1}(i)+\sigma'\xi_{1}(i+j)+\sigma\sigma'\xi_{2}(i,j)$ is the pair probability.  The quantities 
$\xi_{i}$ with $i=1,2$ are given by derivatives of the free energy w.r.t the
inhomogeneous field: $\xi_{1}(i)=<S_{i}^{z}>=\frac{dF}{dh_{i}}$ and
\be
\alpha_{ij}=\xi_{2}(i,j)=<S_{i}^{z}S_{j}^{z}>-<S_{i}^{z}><S_{j}^{z}>
=\frac{d^{2}F}{dh_{i}dh_{j}} \;,
\ee
and the spin susceptibility including SRO contributions is calculated as the
Fourier transform of $\alpha_{ij}$.

In what follows, we will consider only the pair approximation~\cite{sanchez} 
as a first step, so that $(R_{i}-R_{j})=a$, the lattice constant.  To compute 
the thermodynamic response and carrier scattering rate, we need the free 
energy of an effective Ising model in a zeeman field: interestingly, in 
the pair approximation, we show that these can be obtained analytically.  
We find the free energy of the ``core-spin'' subsystem in the pair 
approximation as
\be
F(m,\alpha)=N(-zJ'\frac{m^{2}}{2}-mh_{ext})-zJ'(1-m^{2})\alpha + F[S] \;,
\ee
where
\bn
\nonumber
F[S] &=& k_{B}T \left( \frac{1+m}{2}ln\frac{1+m}{2}
+\frac{1-m}{2}ln\frac{1-m}{2} \right) \\
&+& k_{B}Tz \frac{\alpha^{2}}{4} \;.
\en

The magnetization, $m(T)$, is computed as the first field-derivative of the
free energy:
\be
m(T)=tanh \left[\frac{zJ'm(1-\alpha)+h}{k_{B}T} \right] \;,
\ee
where $\alpha=\alpha_{ij}=[<S_{i}^{z}S_{j}^{z}>-m^{2}(T)]$. Notice the 
difference from the Weiss mean-field (MF) result; the non-local spin 
correlation function explicitly enters the transcendental equation for 
$m(T)$.

The ${\bf q}$-dependent core-spin correlation function is derived from the
second field-derivative of the free-energy:
\be
\chi({\bf q},T)=\frac{-1}{N}
\sum_{i,j}e^{i{\bf q}.{(\bf R_{i}-\bf R_{j})}}[<S_{i}^{z}S_{j}^{z}>-m^{2}(T)]
\ee
\be
\chi({\bf q},T)=
\frac{
[1-\xi_{2}^{2}(T)]}{1+(z-1)\xi_{2}^{2}(T)-2\xi_{2}(T) \varepsilon_{q}^{ij}}
\ee
in $d$ dimensions and 
$\varepsilon_{q}^{ij} \equiv \sum_{a=1}^{d}cos[q_{a}(R_{ia}-R_{ja})]$. In general,
$T_{c}^{FM}$ is determined from the solution of $\chi^{-1}({\bf q},T)=0$ for
$q=0$; i.e, as the solution of $1-(z-1)tanh(J'\beta_{c})=0$.  In $d=1$, this
gives $T_{c}^{FM}=0$, and the correlation length, $\xi(T)=tanh(J'\beta)$, which 
coincides with the result known from the exact solution~\cite{sanchez}.  It 
is easy to see that for small $q$ around ${\bf q}=0$, $\chi({\bf q},T) \simeq
A(T)/(\xi^{-2}+q_{x}^{2}+q_{y}^{2}+q_{z}^{2})$ in $d=3$, which is the classical
Ornstein-Zernicke behavior with the FM spin correlation length given by 
$\xi_{s}^{2}(T)=2tanh(J'\beta)/[1-(z-1)tanh^{2}(J'\beta)]$, again noticeably 
different from the MF result. Comparing with the MF result, we see that 
inclusion of non-local FM spin correlations within the pair-approximation 
(CVM) depresses the transition temperature: $T_{c}^{FM}=T_{c,MF}^{FM}/(z-1)$, 
where $T_{c,MF}^{FM}=2J'z/k_{B}$ is the Weiss MF result, as expected. The $q=0$ 
susceptibility is directly written as,
\be
\chi(0,T)=(g\mu_{B})^{2}\frac{1+tanh(J'\beta)}{1-(z-1)tanh(J'\beta)}
\ee
which deviates noticeably from the MF-Curie-Weiss form at intermediate $T$.

\section{SOME PHYSICAL OBSERVABLES: RESISTIVITY AND THERMODYNAMICS}

Given the spin correlation function, the spin disorder contribution to
the $dc$ resistivity is evaluated using Fermi's golden rule from the
equation~\cite{langer}, extended to an Ising transition.
\be
\rho_{dc}(T)=\frac{m^{*}}{ne^{2}}\int_{FBZ}\chi({\bf q},T)(1-cos\theta)d\theta
\ee
for $D$ dimensions.  Here, $sin(\theta/2)=q/2k_{F}$, whith
$\tau^{-1}(T)=\int_{FBZ}\chi({\bf q},T)(1-cos\theta)d\theta$ again 
involving the non-local spin correlations via $\chi({\bf q},T)$.
We expect the detailed nature of the field induced changes in $\rho_{dc}(T)$ 
to be determined by the field-induced changes in the thermal order parameter
fluctuations. An external field will increase carrier mobility via $t_{ij}$, 
reducing $\tau^{-1}(T)$, and resulting in increased itinerance above $T_{c}$.
 
Calculation of the thermal expansion and magnetovolume effects requires a bit
more work. In a system with Ising-like local moments (magnetic or nematic), 
which are stable as $T$ is raised, one invokes the same localized picture 
as for insulators, where any magnetovolume term is due to the volume 
dependence of inter-site interactions. Following~\cite{campbell}, the 
order parameter fluctuation contribution to the thermal expansion 
coefficient is,
\be
\alpha_{m}(T)=\frac{K}{V}\gamma_{m}\frac{d}{dT}
\left( -\sum_{ij}J[<S_{i}^{z}S_{j}^{z}> -m^{2}(T)] \right) \;,
\ee
where $\gamma_{m}=-dlnJ/dlnV$ is the {\it magnetic} or lattice 
Gr\"uneisen parameter.  The integrated magnetovolume is then,
\be
\frac{\Delta V}{V}=\frac{K}{V}\gamma_{m}\int_{0}^{\infty}dT\frac{d}{dT}
\left( \sum_{ij}J[<S_{i}^{z}S_{j}^{z}>-m^{2}(T)] \right)\;.
\ee

Using Maxwell's thermodynamic relation, the entropy change in a field can be
directly computed from,
\be
\Delta S_{M}(T,h)=S_{M}(T,h)-S_{M}(T,0)=\int_{0}^{B}(\frac{dM}{dT})_{h}dh \;,
\ee
using the equation for the order parameter in the pair approximation 
derived above. (A large change in $\Delta S_{M}(T,h)$ is expected near 
$T_{c}^{FM}$ in DE ferromagnets, as also in disordered local moment 
ferromagnets~\cite{mathur} and would be interesting in the context 
of applications to magnetic refrigeration).
                                                   
The above equations show how short-range order directly influences various 
transport and thermodynamic quantities in metallic systems showing an 
Ising-like semiclassical order via coupling of band carriers to order 
parameter thermal fluctuations. Obviously, our scheme is only valid when 
the Ising-like order arises from microscopic (electronic) degrees of 
freedom that are {\it effectively} decoupled from the band-like carriers, 
but, once established, affects carrier dynamics as shown above.
    
\section{RESULTS}

We now discuss how the above ``simple'' formalism captures, surprisingly,
a wide range of interesting features in the cases (i) and (ii) mentioned in 
the Introduction. 
 
\subsection{Colossal Magnetoresistance Manganites}

\begin{figure}[t]
\includegraphics[width=3.7in]{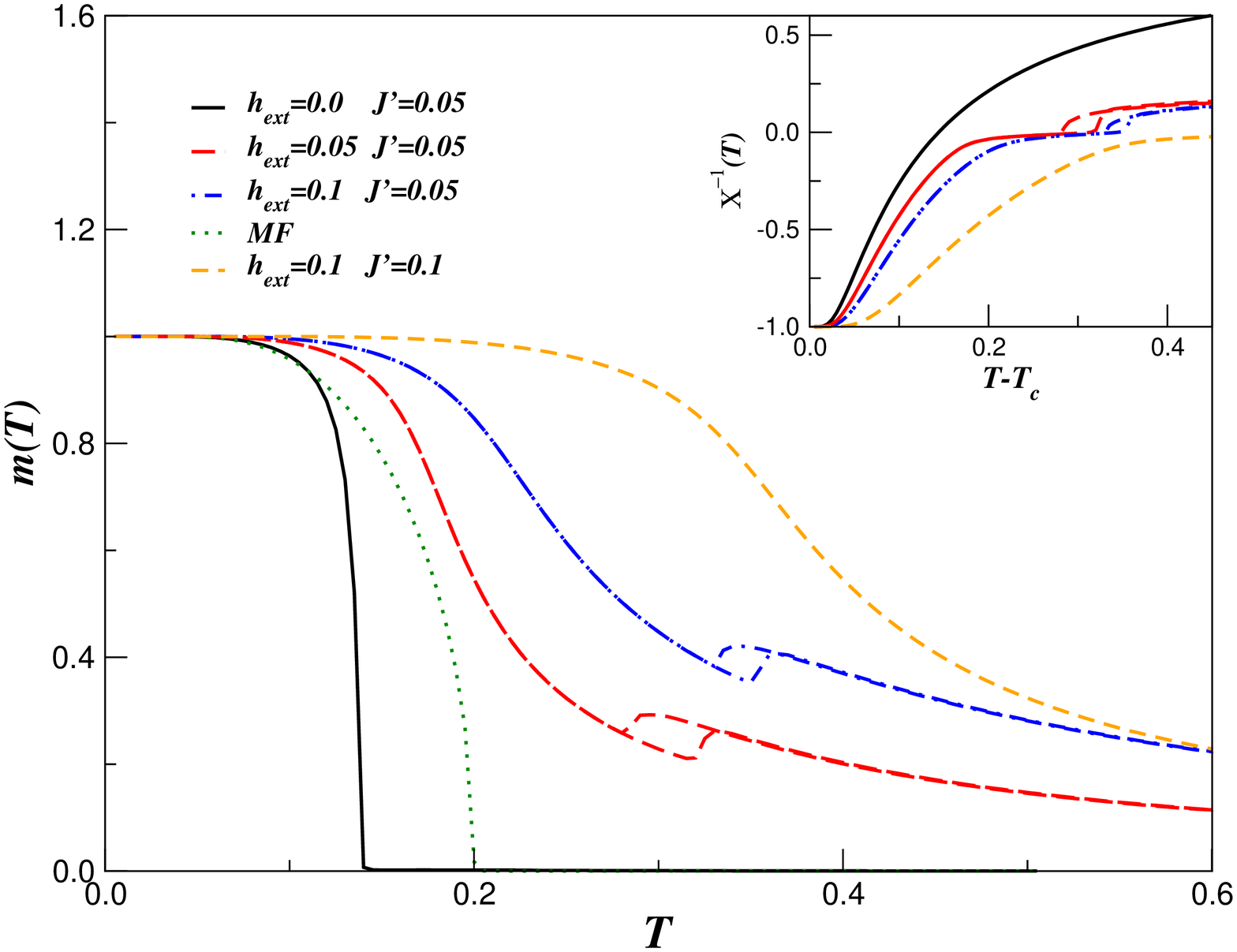}
\caption{
Order parameter as a function of $T$.  The MF $T_{c}$ 
is reduced by 30 percent by order parameter fluctuations.  In a finite 
``field'', clear hysteresis (Ising model in a zeeman field) is seen for 
moderate $h$. In the DE model, a small magnetic field results in appreciable 
FM short-range order above $T_{c}$, modifying thermodynamic and resistive 
behaviors (see Figs~\ref{fig3}-\ref{fig5}). In the nematic case, as in 
underdoped Fe arsenides, the ``field'' is the strain,which acts as a 
conjugate field to the nematicity, i.e, it couples linearly to the nematic 
pseudospin.  Nematic correlations persist well above $T_{T-O}$, in 
qualitative accord with recent indications in underdoped 122-Fe arsenides.}
\label{fig1}
\end{figure}

We note that the Langer-Fisher formulation has been applied with good 
success by Majumdar and Littlewood~\cite{pinaki}.  They considered a variety
of doped magnetic semiconductors.  However, effects of magnetic short-range 
order have received scantier attention in this context, and this is the issue
we will address here.
 
In this section, we show the results of our calculation and discuss them 
in some detail.  In Fig.~\ref{fig1}, we show the core-spin magnetization 
as a function of temperature, $T$.  Notice the difference from the MF 
result, where $m(T)$ vanishes at $T_{c,MF}^{FM}$ (the MF curve is shown 
by the dotted curve in Fig.~\ref{fig1}).  Magnetic SRO ($J'>0$) smears 
the MF transition, as shown in Fig.~\ref{fig1}, the tail showing persistent 
FM SRO above $T_{c}^{FM}$.  A small magnetic field, $h_{ext}=0.05t$, aligns 
these short-range-ordered regions, giving an appreciable $M(T)$ well above 
$T_{c}^{FM}$.  Obviously, pre-existing SRO, in conjunction with a small 
$J'\simeq t^{2}/J_{H}$ makes it easier for a small $h_{ext}$ to polarize 
the high-$T$ phase.  Similar behavior is seen for $h_{ext}=0.1t$.  We 
also clearly see emergence of hysteresis in $m(T)$, a feature expected 
in an Ising model in a zeeman field. The static ($q=0$) spin susceptibility, 
$\chi(0,T)$ shows a distinctly non-Curie-Weiss behavior below 
$T\simeq 2T_{c}^{FM}$, as indeed observed experimentally, showing the 
persistence of magnetic SRO to rather high temperatures.

\begin{figure}[t]
\includegraphics[width=\columnwidth]{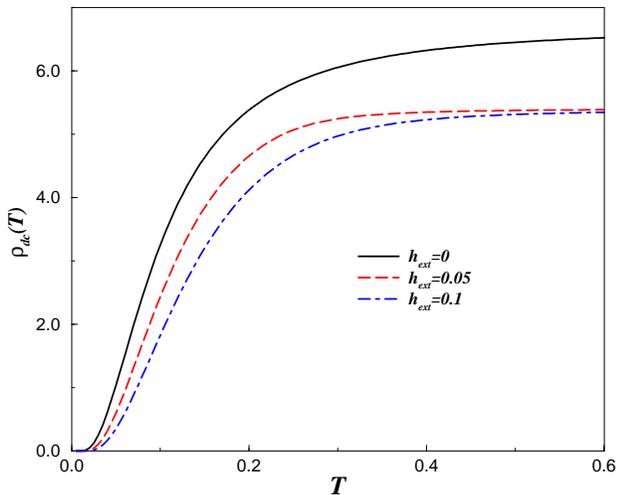}
\caption{
$dc$ resistivity with $J'=0$.  Notice how the resistivity
shows a broad and smmoth change across $T_{c}$.  This is the expected result
for the Ising case, which corresponds to ``potential'' non-magnetic disorder, 
giving $\rho_{dc}(T)\simeq const$ above $T_{c}$.  This is an artifact of the 
MFA and is corrected by the CVM (See Fig.~\ref{fig3}).}
\label{fig2}
\end{figure}

\begin{figure}[t]
\includegraphics[width=3.7in]{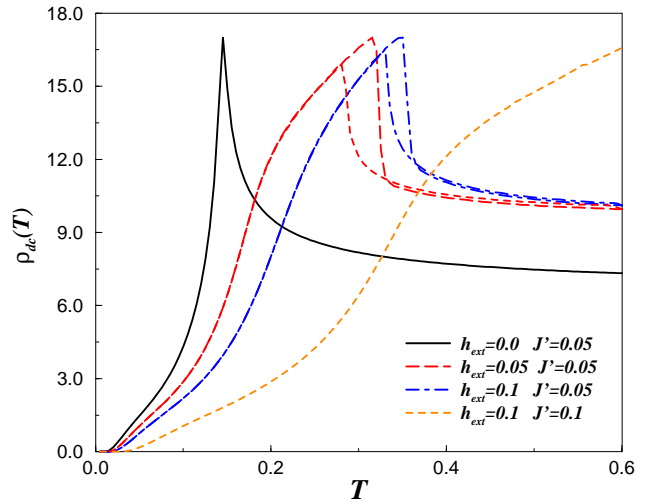}
\caption{
$dc$ resistivity as a function of $T$, with $h=0$ (black
curve) and in a finite ``magnetic'' field (red and blue curves).  Clear
hysteretic behavior for intermediate $h$ is obvious, and is related to that
in $m(T,h)$ vs $h$ in Fig.~\ref{fig1}.  Clear qualitative differences from the
results based upon Weiss mean-field approximation (MFA) (see Fig.~\ref{fig2})
are manifest, and are consequences of the interplay between appreciable 
short-range order and its suppression by an external field.}
\label{fig3}
\end{figure}

The $dc$ resistivity, $\rho_{dc}(T)$, computed within the Born approximation,
actually exhibits an insulator-like behavior above $T_{c}^{FM}$ 
(Fig.~\ref{fig3}). Within simple MF theory,  $\rho_{dc}(T)=const$ above 
$T_{c}^{FM}$ (see Fig.~\ref{fig2} in the case of classical (or Ising) 
spins, since these act like non-magnetic scatterers, giving a resistivity 
characteristic of impurity scattering. With $J'=0$, we thus obtain a result 
resembling that obtained from MF theory, as seen in Fig.~\ref{fig2}. 
With $J'\neq 0$, this behavior is drastically modified: a sharp peak, 
reflecting coupling of carriers to the increasingly singular 
$\chi({\bf q},T)$, is clearly seen in the inset of Fig.~\ref{fig1}. 
We thus conclude that it is precisely the intersite correlations, and, 
in particular, the dominant FM SRO above $T_{c}^{FM}$ which drives the 
insulator-like resistivity in the PM state. The effect of the external 
magnetic field is striking:  $h_{ext}=0.05t$ drives the high T resistivity 
metallic.  This is directly related to the field-dependent suppression 
of the FM short-range correlations, as seen in Fig.~\ref{fig3}, reducing 
the scattering rate, enhancing the carrier mobility and driving the 
system metallic. Similar (enhanced) metallic behavior is seen for 
$h_{ext}=0.1t$, and the dc resistivity shows a huge drop in $\rho_{dc}(T)$ 
as a function of $h_{ext}$. We emphasize that while a large MR drop in 
$h_{ext}$ is plausible within MF treatments of the DE model with additional 
randomness (disordered static JT distortions, static chemical disorder), 
such scenarios require a large disorder (sufficient to split the 
doping-induced impurity band from the lower exchange-split band), 
the origin of which is unclear. Our study should describe 
$Tl_{2}Mn_{2}O_{7}$, where the local moments and band electrons arise 
from distinct orbital states, and do not hybridize much with each 
other~\cite{ventura}.

\begin{figure}[t]
\includegraphics[width=\columnwidth]{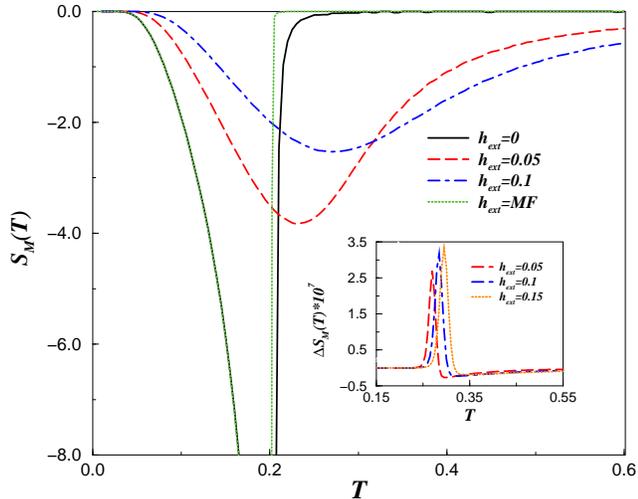}
\caption{
Order parameter fluctuation contribution to the entropy as a
function of $T$.  Significant improvement over the mean-field estimate is
clear.  The field-induced entropy {\it change}, $\Delta S_{M}(T)$, attains 
its maximum in a range of field values where $m(T,h)$ vs $h$ shows hysteretic 
behavior.  For ``local moment'' metallic ferromagnets, combination with a 
coercivity of $O(50)$~Oe would make them attractive candidates for
magnetic cooling applications.  Similar features have recently been 
reported across the T-O structural transition in underdoped Fe arsenides.}
\label{fig4}
\end{figure}

We consider next the magnetic entropy change, computed from the 
$T$-derivative of the magnetization.  In Fig.~\ref{fig4}, we show 
$S_{M}(T,h_{ext})$ for $h_{ext}=0,0.05,0.1$t.  A number of interesting 
features are apparent; the field-induced redistribution of $S_{M}$ is 
drastic, the sharp peak around $T_{c}^{FM}$ changes to a broad peak 
like feature in an external field.  The quantity 
$\Delta S_{M}=[S_{M}(T,H)-S_{M}(T,0)]/S_{M}(T,0)$, which measures field 
induced entropy change shows very encouraging behavior, attaining values 
up to 6 in $h_{ext}=0.15t$.  We also notice that the overall shape of the
curve (shown in the inset of Fig.~\ref{fig4}) is in good qualitative 
agreement with observations~\cite{elbio}. In this context, we notice 
that $Gd$ metal, widely used as a magnetic refrigerant, shows 
$\Delta S_{M}=4.0$ near room temperature.  Thus, $J_{H}>>t$ (small $J$) 
is one of the conditions favoring a large $\Delta S_{M}$, and possible 
application of such systems to magnetic refrigeration.  Further, the 
manganites exhibit considerably small magnetic hysteresis with a coercivity
of 50 Oe near $T_{c}^{FM}$, which should enhance their magnetic cooling 
efficiency.  Finally, interestingly, we remark that the $\Delta S_{M}(T)$ we
extract qualitatively resembles that found for ternary CoMnGe$_{1-x}$Sn$_{x}$
ferromagnetic alloys by Hamer {\it et al.}~\cite{mathur}.  Our formalism, 
based on magnetic SRO, can also be readily extended to include 
structural/chemical SRO in disordered ferromagnetic alloys.  This entails 
incorporation of specific electronic structure details, and lies out of 
scope of the present phenomenological theory.

Finally, we consider the magnetic contribution to the volume expansion, 
$\Delta V_{M}(T)/V$.  Before presenting our results, we present a 
brief physical picture of magnetovolume effects in magnetic systems.  At
high-$T$, with completely disordered local moments, one has as many $\uparrow$
spins as $\downarrow$-spins, so the magnetovolume interactions average to
zero.  As $T$ decreases, short-range local moment correlations begin to 
develop, and the magnetovolume interaction is determined by the spatial 
dependence of the spin correlation function.  Obviously, the MF 
approximation would give incorrect estimates of magnetovolume effects in 
manganites, since, as we have seen, the FM spin correlations are very 
different from those expected from MF approaches.  In view of the ability 
of the CVM to yield a more consistent description of these correlations, 
one expects that it is able to better describe magnetovolume effects in 
such systems.

\begin{figure}[t]
\includegraphics[width=3.7in]{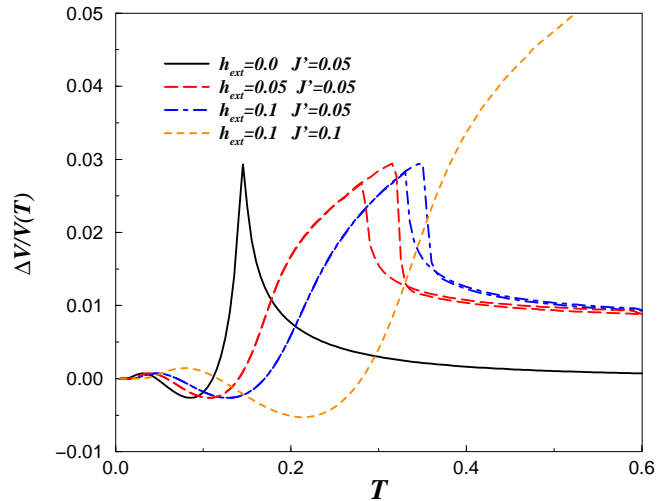}
\caption{
Volume change as a function of $T$.  There is a 
sign change at the ordering temperature ($T_{c}$). Notice the {\it negative} 
sign of $(\Delta V(T)/V)$ below $T_{c}$, followed by a non-linear variation 
and hysteresis above $T_{c}$ when $J'>0$.  Similar features are characteristic 
of some Invar materials.  They have also been reported recently across the 
T-O structural transition in underdoped Fe arsenides.}
\label{fig5}
\end{figure}

In Fig.~\ref{fig5}, we show the magnetovolume changes for different field 
values.  Note that it tends to zero for sufficiently high-$T$, consistent 
with the picture above. Interestingly, at temperatures much higher than 
$T_{c}^{FM}$, the effect of short-ranged FM spin correlations is manifested 
in $\alpha_{M}(T)$, which is {\it negative} and decreases with decreasing 
$T$ upto about $T_{c}^{FM}$. It is interesting to observe that a negative 
$\Delta V_{M}(T)/V$ (thermal expansion coefficient) is characteristic of 
{\it invar} alloys~\cite{invar}, and is observed both in the chemically 
disordered and ordered cases, demonstrating that chemical disorder is 
{\it not} the driving force of invar effects. But thermal spin fluctuations 
introduce spin disorder which could produce such effects.  In our case, the 
magnetovolume changes sign at $T_{c}^{FM}$, so more work is needed to 
describe such systems, which may include cases where there is no 
correlation between $T_{c}^{FM}$ and the sign change in  $\alpha_{M}(T)$.

The field-dependence of $\Delta V_{M}(T)/V$ shows how it is sensitively 
affected by $h_{ext}$.  The fractional change in $\Delta V_{M}(T,h_{ext})/V$ is 
large, and can decrease by as much as 175 percent in $h_{ext}=0.1t$.  
More interestingly, it can be switched in sign by an applied field {\it above}
$T_{c}^{FM}$, as is clearly seen from Fig.~\ref{fig5}. This fact might be 
potentially useful for technological applications.  The change is {\it maximum}
just around $T_{c}^{FM}$, and is thus intimately related to the field-induced
suppression of spin fluctuations above $T_{c}^{FM}$. Hence, it can only be 
accessed in a theory which describes SRO in a consistent way beyond the 
single-site limit. 

\subsection{e-NEMATIC ORDER IN METALS}

Here, we will show that, remarkably enough, our ``simple'' technique also
provides a simple qualitative explanation for a variety of ill-understood
fluctuation effects in physical cases where an electronic nematic 
(e-nematic) instability arises in metals at low $T$.  The whole structure 
of the theory remains intact upon re-identifying the $S^{z}$ as nematic 
pseudospin variables, $N^{z}$.  Moreover the electronic nematic transition 
{\it rigorously} falls into the universality class of an Ising ferromagnet 
in a zeeman field~\cite{carlson}.  Obviously, we want to emphasize that, 
in our effective approach, deeper questions relating to microscopic origin 
of such nematic state(s) {\it cannot} be answered: our main focus is to 
try to understand its {\it consequences} for thermodynamic and transport 
responses. To this end, we will also drastically simplify the more 
complicated situation in real systems of interest 
(like $Sr_{3}Ru_{2}O_{7}$~\cite{arun} and underdoped 
Fe arsenides~\cite{haule,laad,si,vojta,louis}) by replacing their 
undoubtedly strongly correlated (multi) bands by a single free-electron 
like band.  In such systems, the $N^{z}$ may microscopically arise via orbital 
selective Mott localization of a subset of the $d$ orbital band states in the 
real correlated system, and, in our effective model, we {\it assume} that 
this has already occured.  Thus, our analysis {\it cannot} hold for studying
nematic correlations in underdoped cuprates, where the possible nematic order 
and the ``itinerant'' carriers arise from a {\it same} single
band~\cite{sachdev,metzner}.

(i) First, $1/d$ (classical) order parameter fluctuations depress $T_{nem}$ 
by about 30 percent compared to the Weiss MF result.  Without coupling to 
strain, the transition is continuous, but turns into a first-order one with
a finite (not too large) $h_{ext}=\delta(T)$, with hysteretic behavior.  This
is clearly seen in Fig.~\ref{fig3}, where $m(T)$ is now the nematic order
parameter.  The lattice constants will directly react to the onset of nematic 
order, inducing a structural transition (tetragonal-orthorhombic in the 
Fe arsenides).  With reference to underdoped Fe arsenides, this describes the
tetragonal-orthorhombic (T-O) phase transition: this can be either a first- or
second order transition as we traverse the families of real 
Fe arsenides~\cite{phillips}.

(ii) Correspondingly, the $dc$ resistivity shows a clear, sharp anomaly at 
$T_{nem}$ for $\delta=0$, and a ``rounding off'' of this sharp feature for 
a finite $\delta$, with a hysteresis in $\rho_{dc}(T)$ {\it above} $T_{nem}$, 
as shown in Fig.~\ref{fig3}.   

(iii) The entropy change, $\Delta S_{nem}$, now interpreted as the entropy loss
across the e-nematic transition, also exhibits a clear, relatively sharp, 
maximum precisely at $T_{nem}$, as shown in Fig.~\ref{fig4}.  

(iv) Finally, the nematic fluctuation contribution to the thermal expansion 
co-efficient, $\alpha(T)$, shows a clear SRO-induced {\it decrease} above
$T_{nem}$, and a sharp peak at $T_{nem}$, as shown in Fig.~\ref{fig5}.    

Let us now see to what extent these results square up with experimental data
on underdoped Fe arsenides, where recent work on the so-called 122-systems
reveals tantalizing signs of an {\it orbital} nematic state coupled to the 
tetragonal-orthorhombic (T-O) transition.  Surprisingly, we find that effects
of fluctuations of such an e-nematic order (with Ising symmetry) on the 
carriers, reflected in thermodynamic and transport data, seem to be 
qualitatively rationalizable in our ``simple'' model.  

(i) the resistivity, $\rho_{dc}(T)$, shows a maximum at the T-O transition, 
{\it not} at the antiferromagnetic (AF-SDW) transition with ${\bf Q}=(\pi,0)$. 
It also exhibits remarkable hysteretic behavior above $T_{T-O}=150$~K.  This 
goes hand-in-hand with the disappearance of the orthorhombicity, defined as
$\langle O\rangle=\frac{a-b}{a+b}$, where $a,b$ are unit-cell constants (Fe-Fe
nearest neighbor distances) in the FeAs plane.  Interestingly, setting $J'=0$,
i.e, neglecting nematic ordering tendency, kills the prominent peak as well
as the hysteresis (above $T_{T-O}$) in our results, in disagreement with 
experiment (see also below).  If we {\it assume} that a mean field nematic
expectation value, $\langle N^{z}\rangle=(n/2)>0$, develops at the T-O 
transition (though it must be kept in mind that, because the coupling of the
e-nematic order to strain, the e-nematic order cannot be separated from the T-O
distortion, and the e-nematic transition will be smeared), one can address the
issue of the in-plane resistivity anisotropy above the T-O distortion in our
phenomenological approach.  Namely, one can now identify the ``external''
zeeman field value $\delta(T)=h_{z}=n$ with the $N^{z}=1/2$ state with
$\langle O\rangle >0$ and the value $\delta(T)=h_{z}=0$ with the 
$N^{z}=-1/2$ state with $\langle O\rangle <0$ in Ising nematic language. 
From our theoretical resistivity curves (Fig.~\ref{fig3}), we clearly 
see the development of a clear resistivity anisotropy: interestingly, 
it extends to temperatures significantly higher than $T_{T-O}$, and 
achieves its maximum precisely around $T_{T-O}$, where the nematic 
susceptibility is maximal. These findings are broadly consistent with 
experiment~\cite{fisher}, and constitute phenomenological explication 
in terms of an electronic nematic order coupled, by symmetry, to the 
T-O distortion.
 
To the extent that the above e-nematic-plus T-O transition is intimately tied 
to a finite $\langle O\rangle$, its stabilization prepares the ground, via 
anisotropic electronic structure changes, for striped antiferromagnetic 
spin-density-wave (AF-SDW) state with ${\bf Q}=(\pi,0)$ to emerge in a 
natural way.  This program, within the context of a proposal for 
ferro-orbital order, has been carried out in Ref.~\onlinecite{phillips}, 
and the resulting $J_{1a}-J_{1b}-J_{2}$ model with large $J_{2}/J_{1a,b}$
indeed achieves a satisfying description of spin-wave dispersion {\it in} the
AF-SDW state in the O-structure.  Electronic nematicity has hitherto not been
considered within such a program.  In view of the fact that e-nematic and T-O
transitions are strongly coupled (the former is smeared), it follows that 
a similar instability to an AF-SDW state can be worked out in the present case
as well.  We do not do it here, and only mention that, once the T-O distortion
occurs, the scenario of Ref.~\onlinecite{phillips} can take over.  
    
(ii) thermal expansion co-efficient in underdoped Fe arsenides has been 
recently measured by Wang {\it et al.}~\cite{buechner}. Marked anomalies 
in $\alpha(T)$ are found precisely at the T-O transition, and these also 
appear to survive and change with increasing doping: in particular, at 
doping levels close to the T-O boundary at low $T$, the ``fluctuation'' 
contribution, $\Delta\alpha(T)$, becomes {\it negative} below $T_{T-O}$, 
changing sign to positive for $T>T_{T-O}$.  Remarkably, this is exactly
 the form we extract: in our approach (Fig.~\ref{fig5}), it arises due to 
fluctuations associated with Ising-like e-nematic order coupled with the 
T-O distortion.  Adding a purely phenomenological term, linear in $T$, to 
our $\alpha(T)$ computed above could give nice qualitative agreement with these 
experimental results.  Closer inspection, in fact, shows that our result
for $\alpha(T)$ for finite $J'$ (i.e, including nematicity) is in much
closer accord with data than the one with $J'=0$.  In particular, the dip 
(peak) in the (total) measured $\Delta V_{M}(T)/V$ slightly {\it below(above)} 
$T_{T-O}$ is nicely rationalized as arising from short-range fluctuations of an 
Ising-like nematic order associated with ($xz,yz$) orbitals in real underdoped 
Fe arsenides.  The estimated entropy cange across the structural transition
also bears similarities to our computed result, but, since the striped AF 
instability also occurs (slightly below or concurrently with) the structural
instability, a direct comparison between our theory (which only focuses on
nematicity and the accompanying T-O distortion) and experiment is premature.

While the above is by no means a {\it microscopic} description, the 
``localized'' nematic pseudospins ($N^{z}$) can microscopically arise in 
physical situations where sizable multi-orbital electronic correlations 
selectively localize a subset of the relevant planar orbital band states, 
leaving others metallic.  Thus, elucidation of the microscopics of the 
e-nematic-plus T-O transition in Fe-arsenides, and, in particular, 
investigating issues like (i) how such e-nematic-plus T-O transition 
occurs as an instability of the bad-metal ``normal'' 
state~\cite{haule,laad,si,vojta,louis}, and, (ii) its relation (competitor) 
to unconventional superconductivity, involves much more work, and will be 
reported separately.   

Nevertheless, we close this section with a few remarks that may have a 
bearing on recent experimental data for underdoped Fe arsenides.  Given 
that our phenomenolgical Hamiltonian,
\be
H=t\sum_{<i,j>,\sigma}(c_{i\sigma}^{\dag}c_{j\sigma}+h.c) -
J'\sum_{<i,j>}N_{i}^{z}N_{j}^{z} - h_{z}\sum_{i}N_{i}^{z} \;,
\ee 
for the e-nematic transition describes (renormalized in reality) band-like
electrons coupled to a {\it ferro-``magnetic''} Ising model in a zeeman field,
the critical behavior falls into the liquid-gas universality class.  It is 
then perfectly possible that an additional ``tuning parameter'', say static
chemical disorder, could tune the system to the $T=0$ quantum critical 
end-pointof the line of first-order transitions of this quantum liquid-gas 
transition. This could possibly bear a relation to signatures of quantum 
criticality observed in some of the Fe arsenides as a function of doping, 
and would not be inconsistent with the (observed) fact that maximal SC 
$T_{c}(x)$ in Fe arsenides does {\it not} occur at the critical $x=x_{1}$ 
where AF-SDW order vanishes, but rather at $x=x_{2}$ where $T_{T-O}$ would 
have vanished, as is clear by observation of the $T-x$ phase diagrams 
where $T_{T-O}(x)$ and $T_{N}(x)$ are well-separated in $T,x$.  An upshot 
of this reasoning would be then to inquire whether, in a quantum-critical 
scenario, soft fluctuations associated with an underlying QCP associated 
with orbital e-nematic order could act as a pair glue for SC, along lines 
worked out by Si {\it et al.}~\cite{si}.  It would be interesting to 
investigate this line of thinking in more detail, but this is out of 
scope of the present work.

\section{CONCLUSION}

To conclude, in this paper, we have studied how thermally induced SRO and
associated order prameter fluctuations affect various thermodynamic and 
resistive properties in two cases: (i) the double exchange model with 
$J_{H}>>t$, where an effective Ising-like spin model arises, and (ii) an 
electronic system undergoing a phase transition to an e-nematic state, 
coupled to a lattice distortion and shown how a careful treatment of SRO 
goes quite a long way toward a qualitative understanding of several 
striking features in correlated systems of great current interest. Our
analysis should be valid as an {\it effective} phenomenological treatment
in situations where the Ising-like order sets in independently of the 
nature of the band-like electronic state, but, once established, 
drastically affects thermodynamic and transport responses via coupling 
of carriers to the order parameter spin susceptibility. Being extremely 
simple to implement, it can easily be used to analyze experimental 
results in a wide variety of other systems, e.g, in multi-orbital 
systems where orbital order, generically Ising like, plays a crucial 
role in shaping their physical properties.  It can also be used 
very efficiently for Ising models with competing interactions (the 
axial-next-nearest-neighbor Ising (ANNNI) model~\cite{selke}) in a 
zeeman field, which, in itself, is interesting as an effective model 
for complex ordering phenomena in diverse contexts. We plan to address 
such applications in future work.

\acknowledgments
L.C. thanks the Physical Chemistry departement at Technical University 
Dresden for hospitality.


\begin{thebibliography}{28}
     
\bibitem{langer} M.E. Fisher and J.S. Langer, Phys. Rev. Lett. {\bf 20}, 
665 (1968).   

\bibitem{gyorffy} {\it Electrons in disordered metals and at metallic 
surfaces}, edited by P. Phariseau, B.L. Gyorffy, and L. Scheire, New York: 
Plenum Press, (1979). 

\bibitem{kotliar} K. Haule and G. Kotliar, Phys. Rev. B {\bf 76}, 
104509 (2007).
 
\bibitem{furukawa} Y. Motome and N. Furukawa, J. Phys. Soc. Jpn. {\bf 70}, 
1487 (2001).

\bibitem{carlson} E.W. Carlson, K.A. Dahmen, E. Fradkin, and 
S.A. Kivelson, Phys. Rev. Lett. {\bf 96}, 097003 (2006).

\bibitem{arun} S. Raghu, A. Paramekanti, E.-A. Kim, R.A. Borzi, S. Grigera, 
A.P. Mackenzie, and S.A. Kivelson, Phys. Rev. B {\bf 79}, 214402 (2009).  

\bibitem{phillips} W. Lv, T. Kr\"uger, and P. Phillips, Phys. Rev. B 
{\bf 82}, 045125 (2010).

\bibitem{hae} H.-Y. Kee, H. Doh, and T. Grzesiak, J. Phys: Condens. 
Matter {\bf 20}, 255248 (2008).

\bibitem{elbio} {\it Nanoscale Phase Separation and Colossal
Magnetoresistance} by Elbio Dagotto, Springer (2003). 

\bibitem{pinaki} P. Majumdar and P. Littlewood, Nature (London) {\bf 395}, 
479 (1998). Effects of non-local spin correlations involve much more 
sophisticated numerical implementations, for e.g, S. Kumar, A. Kampf, 
and P.Majumdar, Phys. Rev. Lett. {\bf 97}, 176403 (2006). 

\bibitem{fontaine} D. de Fontaine, in {\it Statics and dynamics of 
alloy phase transformations}, Edited by P.E.A. Turchi and A. Gonis, 
North Atlantic Treaty Organization (Scientific Affairs Division) (1992).

\bibitem{sanchez} J.M. Sanchez, Physica A {\bf 111}, 200 (1982).

\bibitem{mathur} J.B.A. Hamer, R. Daou, S. Ozcan, N.D. Mathur, 
D.J. Fray, and K.G. Sandeman, Journal of Magnetism and Magnetic 
Materials {\bf 321}, 3535 (2009).

\bibitem{campbell} I.A. Campbell in {\it Metallic Magnetism}, ed. 
H. Capellmann, Springer (1982).

\bibitem{ventura} C.I. Ventura and B. Alascio, Phys. Rev. B {\bf 56}, 
14533 (1997). 

\bibitem{invar} P. Entel, E. Hoffmann, P. Mohn, K. Schwarz, and V. Moruzzi,
Phys. Rev. B {\bf 47}, 8706 (1993).

\bibitem{haule} K. Haule, H. Shim, and G. Kotliar, Phys. Rev. Lett. 
{\bf 100}, 226402 (2008).

\bibitem{laad} L. Craco, M.S. Laad, S. Leoni, and H. Rosner, 
Phys. Rev. B {\bf 78}, 134511 (2008)

\bibitem{si} J-X. Zhu, R. Yu, H. Wang, L.L. Zhao, M.D. Jones, J. Dai, 
E. Abrahams, E. Morosan, M. Fang, and Q. Si, Phys. Rev. Lett. {\bf 104}, 
216405 (2010), and references therein.  

\bibitem{vojta} A. Hackl and M. Vojta, New J. Phys. {\bf 11} 
055064 (2009).  

\bibitem{louis} N. Doiron-Leyraud, P. Auban-Senzier, S. R. de Cotret, 
C. Bourbonnais, D. J\'erome, K. Bechgaard, and L. Taillefer, 
Phys. Rev. B {\bf 80}, 214531 (2009). 

\bibitem{sachdev} M. Metlitski and S. Sachdev, Phys.Rev. B 
{\bf 82} 075127 (2010).  

\bibitem{metzner} H. Yamase and W. Metzner, Phys. Rev. B 
{\bf 75}, 155117 (2007).  

\bibitem{fisher} J-H. Chu, J. G. Analytis, K. De Greve, P.L. McMahon, 
Z. Islam, Y. Yamamoto, I. R. Fisher, Science {\bf 329}, 824 (2010); 
C. Lester, J-H. Chu, J.G. Analytis, T.G. Perring, I.R. Fisher, and 
S.M. Hayden, Phys. Rev. B {\bf 81}, 064505 (2010).  

\bibitem{buechner} L. Wang, U. K\"ohler, N. Leps, A. Kondrat, M. Nale, 
A. Gasparini, A. de Visser, G. Behr, C. Hess, R. Klingeler, and 
B. B\"uchner, Phys. Rev. B {\bf 80}, 094512 (2009).   

\bibitem{selke} M.E. Fisher and W. Selke, Phys. Rev. Lett. {\bf 44}, 
1502 (1980); W. Selke, Physics Reports {\bf 170}, 213 (1988).

\end{thebibliography}
\end{document}